\documentclass[aps,prb,twocolumn,showpacs,superscriptaddress,floatfix]{revtex4-1}
\usepackage{graphicx,graphics}
\usepackage{dcolumn}
\usepackage{amsmath,amssymb,amsfonts}
\usepackage{latexsym,verbatim}
\usepackage[version=3]{mhchem} 
\usepackage{bm}
\usepackage{color}
\usepackage{ulem}
\usepackage{dsfont}
\usepackage{url}
\usepackage{subfigure}
\usepackage[breaklinks=true,colorlinks,citecolor=blue,linkcolor=blue,urlcolor=blue]{hyperref}

\def\bef{\begin{framed}}
\def\eef{\end{framed}}
\def\be{\begin{equation}}
\def\ee{\end{equation}}
\def\ber{\begin{eqnarray}}
\def\eer{\end{eqnarray}}

\def\nablabold{\mbox{\boldmath $\nabla$}}
\def\sigmabold{\mbox{\boldmath $\sigma$}}
\def\Piv{\mbox{\boldmath $\Pi$}}
\def\rv{{\bm r}}
\def\zv{{\bm z}}
\def\pv{{\bm p}}

\def\vv{{\bm v}}
\def\xv{{\bm x}}
\def\yv{{\bm y}}
\def\jv{{\bm j}}

\def\qv{{\bm q}}
\def\Av{{\bm A}}

\def\fv{{\bm f}}

\def\Mv{{\bm M}}

\def\vv{{\bm v}}
\def\nn{\nonumber}

\begin{document}
\title{Hall viscosity and electromagnetic response of electrons in graphene}
\author{Mohammad Sherafati}
\affiliation{Department of Physics $\&$ Astronomy, University of Missouri, Columbia, Missouri 65211, USA}
\author {Alessandro Principi}
\affiliation{Radboud University, Institute for Molecules and Materials, NL-6525 AJ Nijmegen, The Netherlands}
\author{Giovanni Vignale$^1$}
\email{vignaleg@missouri.edu}
\date{\today}
\begin{abstract}
We derive an analytic expression for the geometric Hall viscosity of non-interacting electrons in a single graphene layer in the presence of a  perpendicular magnetic field. We show that a recently-derived formula in [C. Hoyos and D. T. Son, Phys. Rev. Lett. {\bf 108}, 066805 (2012)], which connects the coefficient of $q^2$ in the wave vector expansion of the Hall conductivity $\sigma_{xy}(q)$ of the two-dimensional electron gas (2DEG) to the Hall viscosity and the orbital diamagnetic susceptibility of that system, continues to hold for graphene -- in spite of the lack of  Galilean invariance -- with a suitable definition of the effective mass. We also show that, for a sufficiently large number of occupied Landau levels in the positive energy sector, the Hall conductivity of electrons in graphene reduces to that of a Galilean-invariant 2DEG with an effective mass given by $\hbar k_F/v_F$ (cyclotron mass). Even in the most demanding case, i.e. when the chemical potential falls between the zero-th and the first Landau level, the cyclotron mass formula gives results accurate to better than 1\%. The connection between the Hall conductivity and the viscosity provides a possible avenue to measure the Hall viscosity in graphene.         
\end{abstract}
\pacs{73.43.Cd, 72.80.Vp, 66.20.Cy, 71.70.Di}
\maketitle
\section{Introduction}
\label{Intro}
Viscosity, i.e. the resistance to a flow in which adjacent parts of a fluid move with different velocities, is a basic property of all classical and quantum liquids, and becomes relevant for electron liquids as well, when disorder and coupling to the lattice are not too strong. As a transport coefficient of a fluid associated with the transport of its momentum, viscosity is a fourth-rank tensor that connects the stress tensor with the rate of change of the strain tensor according to the formula
\begin{align} 
P_{ij}=\sum_{kl}\eta_{ij,kl}v_{kl}\,,
\end{align} 
where $i,j,k,l$ are cartesian indices, $v_{kl}=\frac{1}{2}(\partial_k v_l +\partial_l v_k)$ is the symmetrized gradient of the velocity field $\vv$, and the stress tensor $P_{ij}$ is obtained from the derivative of the Hamiltonian with respect to the metric tensor. In homogeneous rotationally-invariant systems, when the time-reversal symmetry is not broken, the viscosity tensor is entirely described by two scalar transport coefficients, the shear and the bulk viscosities (denoted by $\eta$ and $\zeta$, respectively), which are both dissipative. However, when the tensor is subjected to rigorous scrutiny for a two-dimensional electron gas (2DEG) in a perpendicular magnetic field, it is seen that the broken time reversal symmetry allows, besides the conventional shear and bulk viscosities, the existence of a third {\it non-dissipative} component, known as the Hall viscosity (denoted by $\eta_H$) \cite {Avron95, Avron98} -- also referred to as ``Lorentz shear modulus" \cite{TokatlyVignale07, TokatlyVignale09}. The viscosity tensor in $d=2$ dimensions is then given by
\begin{align}
\eta_{ij,kl}=&\zeta \delta_{ij}\delta_{kl}+\eta(\delta_{ik}\delta_{jl}+\delta_{il}\delta_{jk} - \delta_{ij}\delta_{kl})\nn\\ 
&+\frac{1}{2}\eta_H (\epsilon_{ik}\delta_{jl}+\epsilon_{il}\delta_{jk}+\epsilon_{jk}\delta_{il}+\epsilon_{jl}\delta_{ik})
\end{align}
where $\epsilon_{ij}$ is the rank-2 Levi-Civita tensor.
The last term produces a force density $f_i=-\sum_j \partial_j P_{ij}$ proportional to the Laplacian of the velocity field, but perpendicular to the latter, viz., 
\begin{align}\label{Force}
\fv = \eta_H \nabla^2 \vv \times \hat \zv,
\end{align}
where $\hat \zv$ is the unit vector perpendicular to the plane of the 2DEG. 
The Hall viscosity, $\eta_H = \eta_{xx,xy}$,  is an instance of a class of ``anomalous transport coefficients" -- of which the Hall conductivity  is the best known example -- which are given by the imaginary part of an off-diagonal linear response function, in this case\cite{TokatlyVignale07}
\be\label{KuboFormula}
 \eta_H =\lim_{\omega \to 0} \Im m \frac{\langle\langle P_{xx};P_{xy}\rangle\rangle_\omega}{\omega}\,,
 \ee 
 where $\langle\langle P_{xx};P_{xy}\rangle\rangle_\omega$ is a shorthand for the off-diagonal stress-stress response function. 

Anomalous transport coefficients are invariably expressible in terms of a Berry curvature of the ground-state wave function and can therefore exhibit the striking phenomenon of topological quantization when the Fermi level falls in a spectral gap causing the ordinary dissipative coefficients to vanish. In the present case the Berry curvature involves the derivatives of the wave function with respect to components of the metric tensor \cite{Avron95, TokatlyVignale07}. Indeed, it has been shown \cite{Read09, ReadRezayi11} that in gapped systems such as fractional quantum Hall liquids and $p-$ wave superfluids the ratio between the Hall viscosity and the particle density is determined by the so-called ``shift", a topological quantum number introduced by Wen and Zee \cite{WenZee92}, which arises from the coupling between ``spin" and geometric curvature. Very recently, the Hall viscosity has also been involved in efforts to elucidate the geometric origin of the quantum Hall effect \cite{Haldane15}.

\begin{figure}[!htb]
\includegraphics[angle=0,width=0.5 \linewidth]{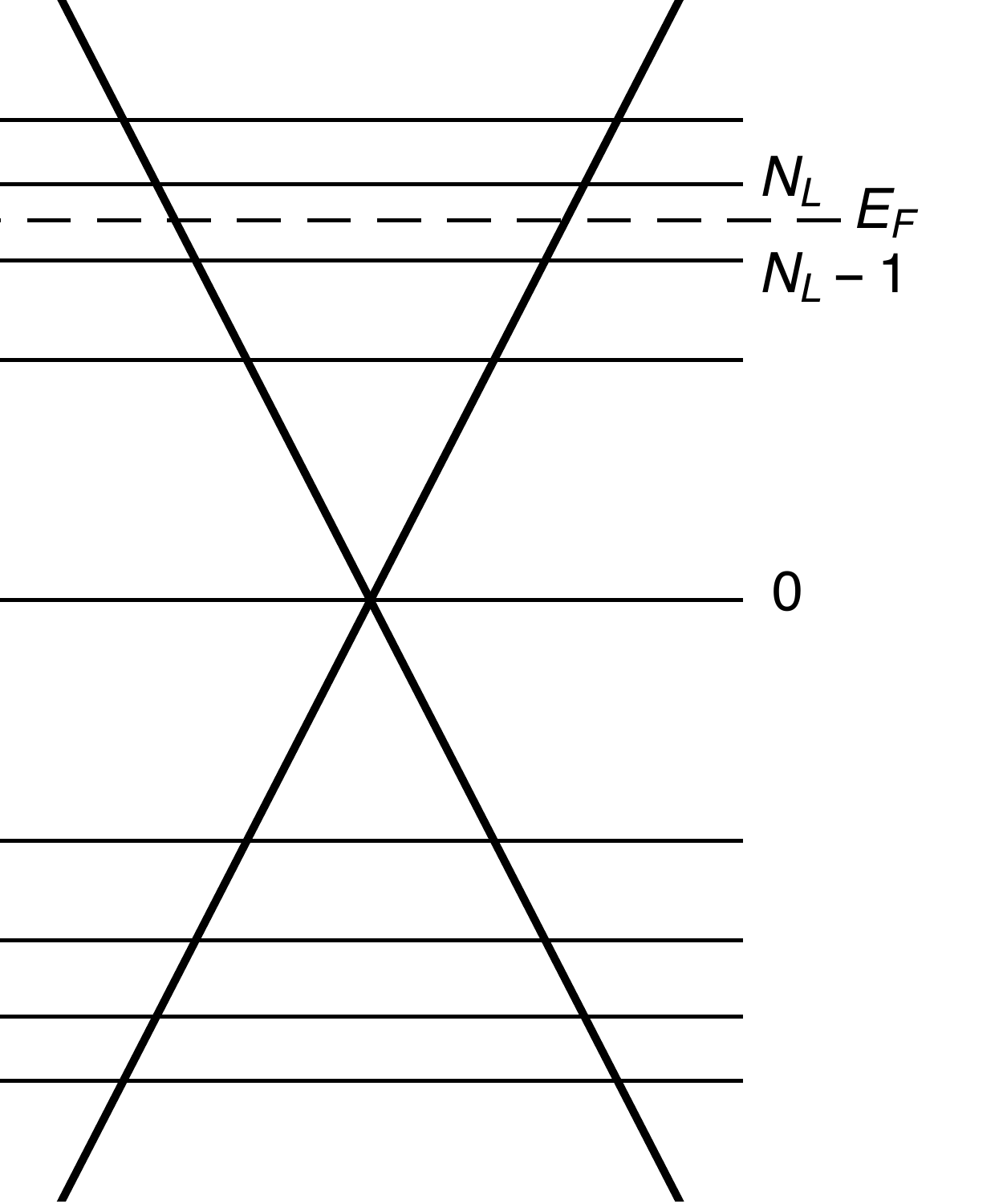}
\caption{Simple schematic cartoon of the Landau levels superposed on the Dirac cone of graphene. The highest occupied and the lowest empty levels are labelled as $N_L-1$ and $N_L$, respectively, compared to the Fermi level which lies within the gap between these two levels. Here for definiteness we assume electron doping, i.e., $N_L\geq 1$.  However, the final formulas will be given in a form that is invariant under the electron-hole transformation $N_L \to -N_L+1$.} 
\label{Fig1}
\end{figure}

In this paper we study the Hall viscosity of electrons in a 2D graphene monolayer~\cite{Neto09} in a perpendicular magnetic field. This system has recently emerged as an excellent candidate for the observation of ordinary viscosity effects \cite{etaGraph}, and we expect it will create similar opportunities for the experimental observation of the Hall viscosity. Just as for the 2DEG, we find that the calculation of the Hall viscosity is meaningful for a perfectly clean non-interacting  electron gas, provided the Fermi level falls in the gap between two Landau levels (LLs).  
Starting from the Kubo formula of Eq.~(\ref{KuboFormula}) one can easily rederive the result of Avron {\it et al.} \cite{Avron95} for the non-interacting 2DEG:
\be\label{etaH2DEG}
\eta_H = g_{sv}\frac{\hbar N_L^2}{8\pi \ell^2}\,,~~~~~{\rm~(2DEG)}
\ee
where $\ell=\sqrt{\hbar c/eB}$ is the magnetic length associated with the magnetic field $B$ with $e$ being the absolute value of the electric charge and $g_{sv}$ is a degeneracy factor taking into account both spins and equivalent valleys ($g_{sv}=2$ for one-valley 2DEG). Here the Fermi level is assumed to fall in the gap between the LLs with indices $N_L-1$ and $N_L$, where the lowest LL has index $0$. Thus, $N_L$ is the filling factor. This formula reduces to that of Ref. [\onlinecite{Avron95}] in the special case of a single full LL, $N_L=1$, with electron density $n=1/(2\pi\ell^2)$. 
We then generalize the above result to the case of electrons in graphene and find
\be\label{etaHGraph}
\eta_H = g_{sv}\frac{\hbar}{4\pi \ell^2}\left[N_L^2+(N_L-1)^2\right]{\rm sgn}\left(N_L-\frac{1}{2}\right)\,,
\ee  
($g_{sv}=4$ for graphene) where, again, the Fermi level falls in a gap between the LLs with indices $N_L-1$ and $N_L$ (see Fig.~\ref{Fig1}). Now, however,  $N_L$ can be zero or negative, corresponding to the possibility of hole doping~\cite{Neto09} and the result exhibits full particle-hole antisymmetry, that is to say, the viscosity changes sign under the transformation $N_L \to -N_L+1$.
Notice that, in the limit of large $N_L$ the Hall viscosity of graphene is four times larger than the Hall viscosity of the 2DEG. We will return to this point below, when we make the connection with the nonlocal Hall conductivity. 

The Hall viscosity calculated in this manner should be called ``geometric Hall viscosity" because it describes the response of the electronic systems to geometric deformation, and it is not a priori related with the ``dynamical Hall viscosity", which should appear in the response of the current density to an external electric field.
In a 2DEG the two definitions coincide, due to the fact that the current density is also the generator of infinitesimal geometrical deformations. Indeed, by expanding the Hall conductivity $\sigma_{xy}(q)$ up to second order in $q$, one finds the Hoyos-Son formula~\cite{Hoyos12}
\be\label{HoyosSon1}
\sigma_{xy}(q)\simeq g_{sv}N_{L} \frac{e^2}{h}\left\{1 + q^2\ell^2\left[\frac{\eta_H}{\hbar n} - N_L\right]\right\}\,.~~~~~{\rm~(2DEG)}
\ee
Both terms in this expression have a simple physical interpretation, which was first presented in Ref.~[\onlinecite{Hoyos12}].  The first term is the Hall current driven by the the viscous force term~(\ref{Force}).  The second term, $-N_L$,  is due to the fact that a shear deformation of the electron gas acts as an effective magnetic field which induces, via the orbital magnetic susceptibility, a non-homogeneous magnetization, hence an additional contribution to the current density.    


In the case of graphene, the dynamical significance of $\eta_H$ can no longer be taken for granted. The difficulty stems from the fact that the velocity operator of electrons in graphene is essentially different from the momentum operator, which would be involved in generating a geometric deformation. The equation of motion for the particle current contains {\it Zitterbewegung} terms \cite{Neto09, Katsnelson06}, which have no analogue in the 2DEG. 

Nevertheless, we have found that the expansion of the $q$-dependent Hall conductivity in graphene can be cast in a form very similar to that of Eq.~(\ref{HoyosSon1}), namely
\begin{align}\label{GrapheneConductivity1}
\sigma_{xy}(q)\simeq g_{sv}&\left(N_{L}-\frac{1}{2}\right) \frac{e^2}{h}\times\nn \\
&\left\{1 + q^2\ell^2\left[\frac{\vert\eta_H\vert}{4\hbar n} - \left\vert N_{L}-\frac{1}{2}\right\vert\right]\right\}\,,
\end{align}
where $n=g_{sv}\frac{|N_L-1/2|}{2\pi \ell^2}$ is the carrier density (electron density for $N_L\geq 1$, hole density for $N_L \leq0$).
In the limit of $N_L \to \infty$ (meaning that the Fermi energy is much larger than the spacing between LLs) the factor $4$ which divides the viscosity in Eq.~(\ref{GrapheneConductivity1}) exactly makes up for the larger viscosity of graphene compared to the 2DEG (compare Eqs.~(\ref{etaHGraph}) and~(\ref{etaH2DEG})). In this limit the formulas~(\ref{HoyosSon1}) and~(\ref{GrapheneConductivity1}) become identical. Furthermore, we will show that the second term in Eq~(\ref{GrapheneConductivity1}) retains the physical interpretation proposed by Hoyos and Son, i.e., can be expressed in terms of the orbital magnetic susceptibility with the appropriate effective mass. The conclusion is that electrons in doped graphene behave (not unexpectedly) like a Galilean-invariant 2DEG with an effective mass $m_c$ which is given by Eq.~(\ref{CyclotronMass}) below.

This paper is organized as follows. In Sec. \ref{model}, we introduce the model Hamiltonian for the non-interacting electrons in graphene in the presence of a perpendicular magnetic field in the continuum limit and the formalism based on the linear-response theory to calculate both the Hall viscosity, presented in Sec. \ref{Hallvisc}, and the conductivity, presented in Sec. \ref{Hallsigma}. Our result for the Hall viscosity of graphene differs significantly from results that have previously appeared in the literature~\cite{Kimura10, Taylor15, Cortijo16}. In Sec. \ref{discussion} we discuss the probable reasons for these differences. Appendixes~\ref{App:ApproxChi} and~\ref{App:ExactChi} present our calculations of the orbital magnetic susceptibility of graphene in a magnetic field.

\section{model Hamiltonian and formalism} 
\label{model}

The effective low-energy Hamiltonian of non-interacting electrons in a monolayer graphene in a perpendicular magnetic field (considering only one spin component and one valley) in a spatially uniform \cite{Footnote1} metrics $g^{ij}$ is
\begin{align}\label{Hamiltonian} 
\hat H = \frac{v_F}{2}\sum \left(\hat \Pi_i g^{ij} \hat \sigma_j+\hat \sigma_i g^{ij} \hat \Pi_j\right)\,,
\end{align} 
where $v_F$ is the Fermi velocity, $\hat \Piv =\hat \pv+\frac{e}{c}\Av(\hat\rv)$ (cgs units will be used throughout the paper) is the kinetic momentum operator, $\Av = By\hat \xv$ is the vector potential corresponding to the magnetic field $B\hat\zv$ in the Landau gauge, $\hat \sigmabold$ is the pseudo-spin operator associated with the two inequivalent sublattices of the honeycomb lattice, $g^{ij}$ is the metric tensor -- which reduces to $\delta_{ij}$ -- in the usual Euclidean geometry, and the sum runs over the particles.  

The stress tensor operator is defined as \cite{TokatlyVignale07} 
\begin{align} \label{Stress}
\hat P_{ij}[g^{ij}]=\frac{2}{\sqrt{g}}\frac{\partial \hat H}{\partial g^{ij}}
\end{align} 
where $g$ is the determinant of the metric tensor. Evaluating the derivative and setting $g^{ij}=\delta_{ij}$ we arrive at the Euclidean stress tensor
\begin{align} \label{Pgraphene}
\hat P_{ij}=v_F\sum \left(\hat \Pi_i\hat \sigma_j+\hat \sigma_i\hat \Pi_j\right)\,.
\end{align} 
From this point on, the metrics will be fixed to Euclidean. The single-particle states are two-component pseudo-spinors of the form \cite{Neto09, Goerbig11}
\begin{align}\label{PsiLL}
|\psi_{\pm nk_y}\rangle=&\frac{1}{\sqrt{2}} \left(\begin{array}{c}|n-1,k_y\rangle\\ \pm|n,k_y\rangle \end{array}\right) \,,~~~(n\geq 1)\, \nn \\
|\psi_{0k_y}\rangle=&\left(\begin{array}{c}0\\ |0,k_y\rangle \end{array}\right),
\end{align}
where $\langle \rv |n,k_y\rangle \propto e^{ik_y y} H_n\left(\frac{x}{\ell}+k_y\ell\right) e^{-\frac{(x+k_y\ell^2)^2}{2\ell^2}}$ are the Landau-gauge wave functions with $H_n(x)$ being the $n$-th order Hermite polynomial and $\ell=\sqrt{\hbar c/eB}$ is the magnetic length associated with the magnetic field $B$. The corresponding energy levels are $E_{\pm n}=\pm \hbar \omega_0\sqrt{n}$, where $\omega_0=\sqrt{2}v_F\ell^{-1}$, and their degeneracy per unit area is $(2\pi\ell^2)^{-1}$. 

We note that for a typical experimentally accessible strength of magnetic fields of $B=10$ T the magnetic length $\ell\approx 257 [\text{\AA}]/\sqrt{B[\text{Tesla}]}$ is approximated to be $\ell\approx 81 \ \text{\AA}\gg a$ where $a\simeq 1.42$ \text{\AA} is the carbon-carbon bond length. Therefore, the present continuum model for graphene electrons in magnetic field is legitimate. In addition, for $B=10$ T, the scaling temperature of the LLs is estimated to be $T=\hbar\omega_0/k_B\approx7100$ K while the Zeeman splitting temperature, $2\mu_BB/k_B\approx13$ K will be relatively negligible. 
Finally, our model for non-interacting electrons excludes all mechanisms that could break the equivalence between the valleys\cite{Goerbig11}.
In other words, each LL has both spin and valley degeneracy, which we denote by $g_{sv}=4$.


\subsection{Calculation of geometric Hall viscosity}
\label{Hallvisc}

To calculate the Hall viscosity we apply the linear-response approach \cite{Kubo}, namely, we use the Kubo formula, Eq.~(\ref{KuboFormula}).
We resort to the Lehmann representation of the Kubo product in the stress-stress response function and express the Hall viscosity (as a function of frequency) in terms of single-particle eigenstates (non-interacting), viz., 
\begin{align}\label{etaH}
\eta_H (\omega)=\frac{2}{\hbar}\frac{1}{2\pi\ell^2}\Im m \sum_{kl}^\prime\frac{[\hat P_{xx}]_{kl}[\hat P_{xy}]_{lk}}{\omega^2-\omega_{lk}^2}\,,
\end{align}
where the prime on the sum means that the energy index $k$ runs over occupied LLs, and the index $l$ runs over unoccupied LLs. The pre-factor accounts for the degeneracy of the LLs. The symbol $[P_{xx}]_{kl}$ denotes the matrix element $[\hat P_{xx}]_{kl}=\langle k|\hat P_{xx}|l\rangle$ and similarly for $[\hat P_{xy}]_{lk}$, and $\omega_{lk}=(E_l-E_k)/\hbar$ denotes the difference of the energies of levels $l$ and $k$. We emphasize that Eq. \eqref{etaH} can be used for both graphene and the 2DEG.  

It is quite simple to apply Eq. \eqref{Stress} to the Hamiltonian for 2DEG ($=(2m)^{-1}\sum\hat \Pi_i g^{ij}\hat\Pi_j$) and plug in the resulting components of the stress tensor into Eq. \eqref{etaH} to recover the well-known expression for the Hall viscosity in $\omega \to 0$ limit \cite{Avron95, TokatlyVignale07}. This was originally obtained by Avron {\it et al.} \cite{Avron95} as a ``Berry curvature" constructed by taking the derivative of the ground-state wave function with respect to the two parameters $g^{xx}$ and $g^{xy}$. The final result is given by Eq.~(\ref{etaH2DEG}).

To actually perform the calculation for graphene it is convenient to express the stress tensor in terms of the dimensionless rising and lowering operators $\hat \Pi_\pm=(2\hbar^2)^{-1/2}\ell(\hat\Pi_x\pm i\hat\Pi_y)$, which satisfy the commutation relation $[\hat \Pi_-,\hat \Pi_+]=1$. From Eq. \eqref{Pgraphene} the respective components of the stress tensor are given by $\hat P_{xx}=2^{-1}\hbar\omega_0(\hat \Pi_+\hat\sigma_+ +\hat \Pi_-\hat\sigma_-)+\hat H$ and $\hat P_{xy}=(2i)^{-1}\hbar\omega_0(\hat \Pi_+\hat \sigma_+-\hat \Pi_-\hat \sigma_-)$, with $\hat \sigma_\pm \equiv \hat \sigma_x\pm i\hat \sigma_y$. Notice that the Hamiltonian operator $\hat H$ in the expression for $\hat P_{xx}$ does not contribute to the calculation of $\eta_H$ since its matrix elements between occupied and unoccupied levels vanish. Plugging these expressions into Eq. \eqref{etaH} and doing some straightforward rearrangements simplifies the equation for graphene to 
\begin{align} \label{etaGraphene}
\eta_H (\omega) = -\frac{\hbar v_F^2}{2 \pi \ell^4}\sum_{kl}^\prime\frac{|[\hat \Pi_+\hat \sigma_+]_{lk}|^2-|[\hat \Pi_-\hat \sigma_-]_{lk}|^2}{\omega^2-\omega_{lk}^2}\,,
\end{align} 
where $\omega_{lk}=\omega_0[\text{sgn}(l)\sqrt{l}-\text{sgn}(k)\sqrt{k}]$. We assume that the Fermi level is in the energy gap between LLs $N_L-1$ and $N_L$ in the positive energy sector and label the unoccupied states with $l \geq N_L$ and the occupied ones with $k\leq N_L-1$, including all the negative energy states; the carrier (electron or hole) density per spin per valley, measured from the neutrality point, will be then given by $n=(2\pi\ell^2)^{-1}|N_L-1/2|$. Notice that the neutrality point is characterized by the chemical potential being set at zero energy so that the zero-th Landau level is half-filled 
and all the negative LLs are completely filled. This explains the shift $1/2$ in the numerator of the carrier density. The Fermi momentum $k_F$ in the presence of a magnetic field can now be easily found from its relationship with the carrier density for the linear bands in pristine graphene, namely, $k_F^2=4\pi n$ (per spin per valley)\cite{Neto09}, which yields $k_F = \ell^{-1}\sqrt{2|N_L-1/2|}$.       

The matrix elements of operators $\hat \Pi_+\hat\sigma_+$ and $\hat \Pi_-\hat\sigma_-$ can be straightforwardly obtained by the following identities:
\begin{align}\label{ImportantFormulas1}
\hat \Pi_+\hat\sigma_+ |\psi_{\pm ik_y}\rangle=&\pm\sqrt{i+1}\left(|\psi_{\pm (i+2)k_y}\rangle + |\psi_{\mp (i+2)k_y}\rangle\right) \nn\\
\hat \Pi_-\hat\sigma_- |\psi_{\pm i k_y}\rangle=&\pm\sqrt{i-1}\left(|\psi_{\pm (i-2)k_y}\rangle -|\psi_{\mp (i-2)k_y}\rangle\right)
\end{align} 
for $i \geq 1$, and 
\begin{align}\label{ImportantFormulas2}
\hat \Pi_+\hat\sigma_+ |\psi_{0k_y}\rangle = \sqrt{2}\left(|\psi_{2k_y}\rangle + |\psi_{-2k_y}\rangle\right),\hat \Pi_-\hat\sigma_- |\psi_{0k_y}\rangle =0\,.
\end{align} 

Proceeding to the evaluation of Eq.~(\ref{etaGraphene}) we observe that a massive cancellation occurs between the contributions of the ``$\Pi_-\sigma_-$ terms" with initial state $k\leq -(N_L+2)$ and those of the ``$\Pi_+\sigma_+$ terms" with $k\leq -N_L$. In the end, all that survives is the contribution of the ``$\Pi_+\sigma_+$ terms", with $k=\pm (N_L-1)$ and $k=\pm (N_L-2)$. Then, making use of Eqs.~(\ref{ImportantFormulas1}) and~(\ref{ImportantFormulas2}) we obtain
\begin{align}\label{etaHGraphOmega}  
\eta_H(\omega) = -\frac{\hbar \omega_0^2}{4\pi \ell^2}\sum_{k=N_L-2}^{N_L-1} \bigg[&\frac{k+1}{\omega^2-\omega_0^2 (\sqrt{k+2}-\sqrt{k})^2}\nn\\ 
&+\frac{k+1}{\omega^2-\omega_0^2(\sqrt{k+2}+\sqrt{k})^2}\bigg].
\end{align}
Setting $\omega=0$ in Eq. \eqref{etaHGraphOmega} and including the spin and valley degeneracy yields Eq.~(\ref{etaHGraph}) quoted in the Introduction.

The above calculation was done under the assumption of positive $N_L$, i.e., the Fermi level was in the positive energy sector. If $N_L$ is negative or zero the Fermi level lies in the negative energy sector and we have hole carriers rather than electron carriers. The calculation proceeds as above, except that the roles of rising and lowering operators, $\hat \Pi_+\hat\sigma_+$ and $\hat \Pi_-\hat\sigma_-$, are now interchanged for both interband and intraband transitions, for example $\hat \Pi_-\hat\sigma_-$ will exclusively be responsible for the intraband transitions within the negative energy sector; this role was played by $\hat \Pi_+\hat\sigma_+$ for transitions within the positive energy sector for $N_L\geqslant 1$. As a result of this interchange, the Hall viscosity [see Eq.~(\ref{etaGraphene})] turns out to have the opposite sign, as indicated by the sign function in Eq.~(\ref{etaHGraph}).
\subsection{Calculation of nonlocal Hall conductivity}
\label{Hallsigma}

In this section, we first present the connection between the Hall viscosity and conductivity for 2DEG and then calculate the transverse Hall conductivity for graphene electrons and demonstrate that a connecting equation similar to that of for the 2DEG exists in the case of graphene.  

By expanding the Hall conductivity $\sigma_{xy}(q)$ of the 2DEG at zero frequency ($\omega=0$) up to second order in $q$, one finds~\cite{Hoyos12} Eq.~(\ref{HoyosSon1}), which can also be rewritten in the physically suggestive form as
\begin{align}\label{HoyosSon2}
\sigma_{xy}(q)\simeq g_{sv}N_L \frac{e^2}{h}\bigg[1 +q^2\ell^2&\bigg(\frac{\eta_H}{\hbar n}+\frac{2\pi}{N_L}\frac{mc^2}{e^2}\chi_d\bigg)\bigg]\,,
\end{align} 
where $\chi_d=-\epsilon''(B)=-\frac{N_L^2}{2\pi}\frac{e^2}{mc^2}$ is the negative of the second derivative of the energy density $\epsilon(B)=\frac{e^2N_L^2B^2}{4\pi m c^2}$, with respect to $B$, {\it taken at constant filling factor}.  We note that the latter specification -- that the Fermi level remain locally in the gap between two LLs -- is essential to obtaining the correct formula. From this point of view, our orbital susceptibility $\chi_d$ is non-standard, because we are changing the electron density as well as the magnetic field so as to keep the ratio $n/B$ ($\propto N_L$) constant. Because $\sigma_{xy}(q)$ controls the response of the current to an external electric field, we see that the geometric Hall viscosity $\eta_H$ is also the dynamical Hall viscosity. 
A simple physical interpretation of  Eq. \eqref{HoyosSon1} was first presented in Ref.~[\onlinecite{Hoyos12}].  The zero-th order term $\sigma_{xy}(0)=N_L \frac{e^2}{h}$ is the celebrated universal Hall conductivity for a 2DEG. The first term is the Hall current driven by the the viscous force term given in Eq. \eqref{Force}. The second term is due to the fact that a shear deformation of the electron gas acts as an effective magnetic field which induces, via the orbital magnetic susceptibility, a non-homogeneous magnetization $\Mv(\rv)$, which contributes to the current  density via the formula $\jv(\rv) = \nablabold \times M(\rv)$. 


As discussed in the Introduction, the dynamical significance of $\eta_H$ can no longer be taken for granted in graphene.  
The nonlocal Hall conductivity $\sigma_{xy}(q,\omega)$ must be calculated from the Kubo product $\langle\langle \hat j_x(\qv);\hat j_y(-\qv)\rangle\rangle_\omega$ where the current density operator is given by
\begin{align} 
\hat \jv(\qv) = -ev_F \sum \hat \sigmabold e^{-i\qv\cdot \hat\rv}
\end{align} 
with the sum running over the particles. Making use of this expression in the standard formulas of linear response theory we easily arrive at
\begin{align} \label{KuboConductivity}
\sigma_{xy}(q,\omega)=\frac{e^2}{h} \omega_0^2 \Im m \sum_{kl}^\prime \frac{[\hat\sigma_xe^{-iqy}]_{kl}[e^{iqy}\hat \sigma_y]_{lk}}{\omega^2-\omega_{lk}^2}
\end{align} 
where we have oriented the wave vector along the $y$ direction, $\qv =q\hat\yv$, to take full advantage of the Landau gauge. Notice that, due of the presence of the $e^{\pm iqy}$ the matrix elements now involve states with different values of $k_y$, namely $|\psi_{k, k_y}\rangle$ and $|\psi_{l,k_y+q}\rangle$. Keeping in mind that the shift $k_y \to k_y+q$ corresponds to a shift of the center of the wave function by $q \ell^2\hat \xv$, and that the shift operator is $\frac{d}{dx} = \frac{i}{\hbar}\hat \Pi_x$, we easily arrive at the following expansion of the matrix elements to order $q^2\ell^2$:
\begin{align}
\left[\hat \sigma_xe^{-iqy}\right]_{kl} &=&\langle \psi_{k,k_y}|\hat \sigma_x \left(1+ i \bar q\bar \Pi_x -\frac{1}{2}\bar q^2 \bar \Pi_x^2\right)|\psi_{l,k_y}\rangle\,, \nn\\
\left[e^{iqy}\hat\sigma_y\right]_{lk} &=&\langle \psi_{l,k_y}|\left(1- i \bar q\bar \Pi_x -\frac{1}{2}\bar q^2 \bar \Pi_x^2\right)\hat \sigma_y|\psi_{k,k_y}\rangle\,,
\end{align}
where $\bar q =q\ell$ and $\bar \Piv = \frac{\ell}{\hbar}\hat \Piv $. Substituting these expansion in Eq.~(\ref{KuboConductivity}) and keeping terms up to order $q^2$ we obtain
\begin{align}\label{SigmaXY}
\sigma_{xy}(q,&\omega)=\frac{e^2}{h} \omega_0^2 \Im m \sum_{kl}^\prime \frac{[\hat\sigma_x]_{kl}[\hat \sigma_y]_{lk}}{\omega^2-\omega_{lk}^2}\nn\\
&+\frac{e^2}{h} \bar q^2 \omega_0^2 \Im m \sum_{kl}^\prime \frac{[\hat\sigma_x \bar \Pi_x]_{kl}[\hat \sigma_y\bar\Pi_x]_{lk}}{\omega^2-\omega_{lk}^2}\nn\\
&-\frac{e^2}{h} \frac{\bar q^2}{2} \omega_0^2\Im m \sum_{kl}^\prime \frac{[\hat\sigma_x]_{kl}[\hat \sigma_y\bar\Pi_x^2]_{lk}+[\hat\sigma_x\bar \Pi_x^2]_{kl}[\hat \sigma_y]_{lk}}{\omega^2-\omega_{lk}^2}\,.
\end{align}
The first term on the right-hand side of Eq. \eqref{SigmaXY} is best evaluated by replacing $\hat \sigma_x = 2^{-1}(\hat \sigma_+ +\hat \sigma_-)$, $\hat \sigma_y =(2i)^{-1}(\hat \sigma_+ -\hat \sigma_-)$, and then making use of formulas such as
\begin{align}
\hat \sigma_\pm |\psi_{\pm k,k_y}\rangle =&\pm \left(|\psi_{\pm(k\pm1),k_y}\rangle\pm |\psi_{\mp(k\pm1),k_y}\rangle  \right)\,,~~~~k\geq 1\nn\\
\hat \sigma_+ |\psi_{0,k_y}\rangle=&\sqrt{2}\left(|\psi_{1,k_y}\rangle+|\psi_{-1,k_y}\rangle\right)\,,~~\hat \sigma_- |\psi_{0,k_y}\rangle = 0\,. 
\end{align}  
After a massive cancellation of contributions from negative energy states only the contributions from $k= N_L-1$ and $k=-N_L-1$ survive yielding 
\begin{align} 
\sigma_{xy}(q=0,\omega)=-\frac{e^2}{h}&\frac{\omega_0^2}{4}\bigg[\frac{1}{\omega^2-\omega_0^2(\sqrt{N_L}-\sqrt{N_L-1})^2}\nn\\
&+\frac{1}{\omega^2-\omega_0^2(\sqrt{N_L}+\sqrt{N_L-1})^2}\bigg].
\end{align} 
Setting $\omega=0$ we immediately recover the well-known expression for the ``anomalous" Hall conductivity of electrons in graphene \cite{AQHGraph}
\begin{align} 
\sigma_{xy}(q=0,\omega=0)=\frac{e^2}{h}\left(N_L-\frac{1}{2}\right)\,.
\end{align} 
We note that this well-established result could not have been obtained without proper inclusion of the negative energy states.

The second term on the right-hand side of Eq.~(\ref{SigmaXY}) is readily expressed in terms of the Hall viscosity, after noting that $[\hat \sigma_x \bar \Pi_x]_{kl}=(\hbar\omega_0\sqrt{2})^{-1}[\hat P_{xx}]_{kl}$ and $[\hat \sigma_y \bar \Pi_x]_{lk}=\ell (\hbar\omega_0\sqrt{2})^{-1}[\hat P_{yx}]_{lk}$. Comparing the resulting expression with Eq.~(\ref{etaH}) for the Hall viscosity and noting that $n=(2\pi\ell^2)^{-1}(N_L-1/2)$ (per valley per spin for electrons) we see that the second term on the right-hand side of Eq.~(\ref{SigmaXY}) can be written as
\begin{align} \label{sigmaxy2a}
\sigma_{xy}^{(2a)}(q,\omega) = \frac{e^2}{h}\left(N_L-\frac{1}{2}\right)\frac{\eta_H(\omega)}{4\hbar n} (q\ell)^2\,.
\end{align}  

Lastly, consider the third term on the right-hand side of Eq.~(\ref{SigmaXY}) and replace $\bar \Pi_x^2 = \frac{1}{2}\left(\hat \Pi_+^2+\hat \Pi_-^2+2\hat \Pi_+\hat\Pi_-  +1\right)$. It is easy to verify that only the $\hat \Pi_+ \hat \Pi_-$ terms can contribute and that
\begin{align} 
 \hat \Pi_+ \hat \Pi_- |\psi_{\pm k,k_y}\rangle &=\left(k-\frac{1}{2}\right)|\psi_{\pm k,k_y}\rangle -\frac{1}{2} |\psi_{\mp k,k_y}\rangle\,, \nn\\
 \hat \Pi_+ \hat \Pi_- |\psi_{0,k_y}\rangle &=0.
 \end{align} 
Again, a massive cancellation of contributions from negative energy states takes place, after which only the contributions from $k= N_L-1$ and $k=-N_L-1$ survive yielding
 \begin{align} \label{sigmaxy2b}
\sigma_{xy}^{(2b)}(q,\omega) = \frac{e^2}{h}&\frac{\omega_0^2}{4}\bigg[\frac{N_L-\frac{1}{2}}{\omega^2-\omega_0^2(\sqrt{N_L}-\sqrt{N_L-1})^2}\nn\\
&+\frac{N_L-\frac{1}{2}}{\omega^2-\omega_0^2(\sqrt{N_L}+\sqrt{N_L-1})^2}\bigg](q\ell)^2.
\end{align}  
Setting $\omega=0$, reinstating the degeneracy factor $g_{sv}$ and preserving the electron-hole symmetry (as $N_L\to -N_L+1$ then $\sigma_{xy}(q) \to -\sigma_{xy}(q)$) we obtain the connecting equation between the Hall conductivity and the Hall viscosity for graphene similar to the one for the 2DEG [Eq. \eqref{HoyosSon1}], viz.,  Eq.~(\ref{GrapheneConductivity1}).
Furthermore, we find that the in the limit of $N_L\to \infty$ the second term in the square brackets of Eq.~(\ref{GrapheneConductivity1}) can be rewritten in a form similar to that of  Eq.~(\ref{HoyosSon2}), namely
\begin{widetext}
\begin{align} \label{GrapheneConductivity2}
\sigma_{xy}(q)\simeq &g_{sv}\left(N_{L}-\frac{1}{2}\right)\frac{e^2}{h}\bigg[1 + q^2\ell^2\bigg(\frac{\vert\eta_H\vert}{4\hbar n}+\frac{2\pi}{|N_L-1/2|}\frac{m_cc^2}{e^2}\chi_d\bigg)\bigg],{\rm~~~(Graphene)}
\end{align} 
\end{widetext}
where $\chi_d$ is the orbital magnetic susceptibility, which we calculate in Appendixes (\ref{App:ApproxChi}) and (\ref{App:ExactChi}).
For large $N_L$ it is given by [ See Appendix (\ref{App:ApproxChi})] 
\begin{align} \label{OrbitalMagneticSusceptibility}
\chi_d=-\frac{e^2\sqrt{2}}{4\pi \hbar c}\frac{v_F \ell}{c}\left[\left\vert N_L-\frac{1}{2}\right\vert^{3/2}-\frac{1}{32}\left\vert N_L-\frac{1}{2}\right\vert^{-1/2}\right]
\end{align} 
and $m_c$ is the ``cyclotron mass" at this limit given by [See Appendix (\ref{App:ApproxChi})]
\begin{align} \label{CyclotronMass}
m_c \equiv \frac{\hbar k_F}{v_F}=\frac{\hbar \sqrt{2|N_L-1/2|}}{v_F\ell}.
\end{align} 

In the limit of $N_L \to \infty$ (meaning that the Fermi energy is much larger than the spacing between LLs) the factor $4$ which divides the viscosity in Eq.~(\ref{GrapheneConductivity1}) exactly makes up for the larger viscosity of graphene compared to the 2DEG. In this limit, the formulas~(\ref{HoyosSon2}) and~(\ref{GrapheneConductivity2}), as well as (\ref{HoyosSon1}) and ~(\ref{GrapheneConductivity1}), become identical. Electrons in graphene behave (not unexpectedly) like a Galilean-invariant 2DEG with an effective mass $m_c$ given by Eq.~(\ref{CyclotronMass}). Interestingly, from the exact evaluation of the orbital magnetic susceptibility (see Appendix \ref{App:ExactChi}) it turns out that the above statements are essentially exact even for the smallest values of $N_L$, e.g.,  $N_L=1$. 

\section{Discussion}
\label{discussion}

In this section we compare our result for the geometric Hall viscosity in graphene with those obtained earlier in the literature and address the origins of the discrepancies in the results and shortly comment on the difference between the methods used. 

Our results for the geometric Hall viscosity in graphene [Eq. \eqref{etaHGraph}] differ significantly from those recently obtained by Cortijo {\it et al.} \cite{Cortijo16}, where a Hall viscosity was obtained from the response of the electron liquid to a deformation of the graphene lattice: as a result, their Hall viscosity is proportional to the Gr\"uneisen parameter -- a property of the graphene lattice that connects the lattice strain to a pseudo-magnetic field experienced by the electrons \cite{Vozmediano10}. In our work the geometric deformation is applied to the electron liquid in the continuum limit, i.e., on a length scale much larger than the lattice constant.  This kind of deformation is naturally created by slowly varying electric fields, which arise in hydrodynamic flow, optical excitations, and plasmons. Any lattice property, such as the  Gr\"uneisen parameter, becomes irrelevant in this limit.  It is for this reason that our Hamiltonian retains the simple form of Eq.~(\ref{Hamiltonian}): additional terms that would appear if the metric tensor were allowed to be position dependent \cite{Vozmediano10} are negligible in the continuum limit.

In addition, there exists two earlier studies in which the contribution of an individual Landau level indexed $n$ into the geometric Hall viscosity of graphene denoted by $\eta^{(n)}_H$ has been reported: Kimura \cite{Kimura10} arrives at his results from the calculation of the Berry curvature for the Dirac fermions adapted from the result for 2DEG LLs \cite{Levay} while Tuegel and Hughes~\cite{Taylor15} apply momentum-transfer method to a continuum model for Dirac fermions and from the connection between Schr\"odinger Landau levels and the Dirac ones obtain their results. Tuegel and Hughes report that this contribution is given by $\eta^{(0)}_H=\hbar/\left(8\pi \ell^2\right)$ for $n=0$ and $\eta^{(n)}_H=\hbar \left|n\right| /\left(4\pi \ell^2\right)$ for $n\neq 0$ \cite{Taylor15}. 

Negative values of $n$ refer to LLs filled with holes. Both studies agree for $n\neq0$ but for the zero-th Landau level Kimura's result is half of the value reported in Ref. [\onlinecite{Taylor15}]. Our result, according to Eq. \eqref{etaHGraph},
is four times larger than in Ref. [\onlinecite{Taylor15}]:
\begin{align}\label{OurResult}
  \eta_H^{(n)} =
  \begin{cases}
    \hbar / \left(2\pi \ell^2\right)
    & n = 0
    \\
    \hbar \left|n\right| / \left(\pi \ell^2\right)
    & n \neq 0.
  \end{cases}
\end{align}

We believe that the extra factor arises from the inclusion of the negative energy states. In Ref. [\onlinecite{Taylor15}], Section V,  the contribution of the $n$-th Landau level to the Hall viscosity  was adapted from the corresponding result   obtained from the deformation of the  ``Schr\"odinger Landau levels".  But the  ground state of massless graphene, in the continuum approximation, consists of an infinite number of occupied negative-energy LLs. 
The deformation of a positive-energy Landau level $n$ creates a superposition that includes a negative-energy Landau level $-(n+2)$. This can be seen clearly in Eqs. \eqref{ImportantFormulas1} and \eqref{ImportantFormulas2}, where one can see that the stress tensor, acting on level $i$ changes it into a superposition of levels with $i+2$ and $-(i+2)$. We believe that the contribution of the negative-energy states to the deformation of the wave function creates the difference between the results for Schr\"odinger and Dirac Landau levels. Curiously, the extra factor $4$ that we obtain from the inclusion of the negative energy states is ultimately cancelled by another factor $4$ in the final expression for the conductivity, Eq.~(\ref{GrapheneConductivity2}). As discussed after Eq.~(\ref{GrapheneConductivity2}), this is the factor that effectively restores the Galilean invariant form of the relation between  conductivity and viscosity. One may say that $\eta_H/4$ is the ``dynamical" Hall viscosity, as opposed to the ``geometric" Hall viscosity.

We conclude this section with a few words on the different methods used to calculate the Hall viscosity. For the non-interacting case, definitely, the linear-response method we have used in this paper dramatically facilitates the calculation of the Hall viscosity compared to the much more complicated methods based on the Berry curvature \cite{Avron95, Read09} or effective-action theories \cite{Hoyos12}. However, the situation may change when electron-electron interaction are of essence, e.g., in the case of the fractional quantum Hall liquid. In this case the single-particle picture fails and  the momentum-transfer method used in Ref. [\onlinecite{Taylor15}] or the general effective-action method may serve the purpose more suitably.  

\section{Summary}
\label{summary}

From the stress-stress linear response function, we obtain an analytical expression for the geometric Hall viscosity of electrons in monolayer graphene in the presence of a perpendicular magnetic field. We then demonstrate that although graphene is not a Galilean-invariant system, the connecting equation between the Hall conductivity and viscosity first derived by Hoyos and Son \cite{Hoyos12} for Galilean-invariant systems still holds for graphene provided that the effective cyclotron mass is properly defined for large number of Landau levels lying within the linear bands. Since the nonlocal conductivity directly controls the electrodynamic response of the electron liquid, the existence of such a deep connection between viscosity and conductivity creates a very real possibility of accessing the Hall viscosity of graphene from conductivity measurements.   

\section*{Acknowledgements}
The work at the University of Missouri was supported by DOE Grant No. DE- FG02-05ER46203. A.P. acknowledges support from ERC Advanced Grant 338957 FEMTO/NANO. We thank Ilya Tokatly and Rudro Rana Biswas for enlightening discussions. We especially thank Thomas Tuegel and Taylor Hughes for helping us to correct an error in an earlier version of this paper.
\onecolumngrid
\appendix
\section{Evaluation of the orbital magnetic susceptibility of graphene}
\label{AppA}

In this Appendix, we calculate the ground-state energy density and its second derivative with respect to the magnetic field, that is to say, the orbital diamagnetic susceptibility. First, we arrive at the approximate results using the Euler-McLauren formula and then we obtain the exact results with the aid of the Ramanujan formula for the sum of the square roots of integers.

\subsection{{\textit{Approximate}} evaluation of the orbital magnetic susceptibility}
\label{App:ApproxChi}

The zero-magnetic field and zero-temperature orbital magnetic susceptibility of the electron gas in graphene is known to vanish~\cite{McClure, Principi10} at finite density, and to become infinitely diamagnetic (i.e. negative) at the neutrality point. Here we calculate the zero-temperature orbital magnetic susceptibility for the incompressible non-interacting liquid at finite magnetic field, and show that it is connected to the $\sigma_{xy}^{(2b)}$ term [Eq.~(\ref{sigmaxy2b})] in the expansion of the conductivity, in a manner that is very similar to a 2DEG, provided a suitable effective mass is defined.

The orbital magnetic susceptibility is defined as $\chi_d = -\epsilon''(B)$, where $\epsilon(B)$ is the energy density. As we discussed earlier, it is essential that the second derivative with respect to magnetic field be taken at constant filling factor. Let us begin by writing the expression for $\epsilon(B)$ when the Fermi level is between Landau levels $N_L-1$ and $N_L$ in the positive energy sector ($N_L\geq 1$):
\begin{align}\label{EpsB} 
\epsilon(B)&=-\frac{\hbar v_F\sqrt{2}}{2\pi \ell^3}\sum_{k=1}^\infty \sqrt{k}+\frac{\hbar v_F\sqrt{2}}{2\pi\ell^3}\sum_{k=1}^{N_L-1}\sqrt{k}\nn \\
&=-\frac{\hbar v_F\sqrt{2}}{2\pi\ell^3}\sum_{k=N_L}^\infty \sqrt{k}\,.
\end{align} 
We note that the same expression works also for the negative energy sector, i.e., for $N_L\leq0$, after performing the electron-hole transformation $N_L \to -N_L+1$. Rather than summing to infinity we sum to a cutoff $N_c$ determined by the condition
\begin{align}\label{Nc} 
\frac{v_F\sqrt{2}}{\ell}\sqrt{N_c}=v_F k_c  \rightarrow N_c = \frac {k_c^2\ell^2}{2}
\end{align} 
where $k_c$ is a cutoff in momentum space. For magnetic fields of the order $B=10$ T and using $\ell\approx 257 [\text{\AA}]/\sqrt{B[\text{Tesla}]}$ and the ultraviolet momentum cut-off for the linear bands in graphene $k_c\sim 1/a$ \cite{deJuan10} with $a\approx1.42$ \AA \, Eq. \eqref{Nc} gives $N_c\approx1600$ LLs lie within the linear bands. 

Using the cut-off level $N_c$ and Eq. \eqref{Nc}, Eq. \eqref{EpsB} can be written as 
\begin{align} \label{EnergyDensity}
\epsilon(B)=-\frac{\hbar v_F\sqrt{2}}{2\pi\ell^3}\sum_{k=N_L}^{\frac {k_c^2\ell^2}{2}} \sqrt{k}
\end{align} 
Next we use the Euler-McLaurin formula \cite{Abramowitz72} to approximate the sum for large $N_c$:
\begin{align}\label{EulerMcLaurin}
\sum_{k=N_L}^{\frac {k_c^2\ell^2}{2}} \sqrt{k}&\simeq \int_{N_L}^ {\frac {k_c^2\ell^2}{2}} dx \sqrt{x}+\frac{1}{2}\left(\sqrt{N_L}+\frac{k_c\ell}{\sqrt{2}}\right)+\frac{B_2}{2!}\left(\frac{1}{2\sqrt{\frac {k_c^2\ell^2}{2}}}-\frac{1}{2\sqrt{N_L}}\right)+\mathcal{O}[(k_c^2\ell^2)^{-5/2}-N_L^{-5/2}]\nn\\
&\simeq\frac{k_c^3\ell^3}{3\sqrt{2}}+\frac{k_c\ell}{2\sqrt{2}}-\frac{2}{3}N_L^{3/2}+\frac{1}{2}N_L^{1/2}-\frac{1}{24}N_L^{-1/2}\,,
\end{align}
in which we have used the Bernoulli number $B_2=1/6$ and that the higher order terms in $k_c$ and $N_L$ are negligible as $k_c\to \infty$ in comparison with the leading terms proportional to $k_c^3\ell^3$, $k_c\ell$, $N_L^{3/2}$ and $N_L^{1/2}$, respectively. Working at this order of accuracy, for large $N_L$, we can rewrite the last three terms on the right hand side of Eq. \eqref{EulerMcLaurin} as
\begin{align}
-\frac{2}{3}N_L^{3/2}+\frac{1}{2}N_L^{1/2}-\frac{1}{24}N_L^{-1/2} = -\frac{2}{3}\left\vert N_L-\frac{1}{2}\right\vert^{3/2}+\frac{1}{48}\left\vert N_L-\frac{1}{2}\right\vert^{-1/2}
\end{align} 
which exhibits the exact electron-hole symmetry. The first and the second term on the right-hand side of Eq. \eqref{EulerMcLaurin}, when used in the expression for $\epsilon(B)$ become independent of $B$ and linear in $B$, respectively. Because their second derivative with respect to $B$ vanishes these terms, although formally divergent, do not contribute to the orbital magnetic susceptibility. The remaining part in is proportional to $B^{3/2}$, viz.,
\begin{align}\label{AproxEps} 
\epsilon(B) \simeq  \frac{\hbar v_F\sqrt{2}}{3\pi\ell^3}\left(\left\vert N_L-\frac{1}{2}\right\vert^{3/2}-\frac{1}{32}\left\vert N_L-\frac{1}{2}\right\vert^{-1/2}\right)
\end{align} 
Taking the second derivative of Eq. \eqref{AproxEps} with respect to $B$ we arrive at the orbital magnetic susceptibility, $\chi_d$, expressed in Eq. \eqref{OrbitalMagneticSusceptibility} in the main text.

We now plug the resulting $\chi_d$ into Eq.~(\ref{GrapheneConductivity2}) and we wish to define the effective mass $m_c$ in such a way that the result of the microscopic calculation [Eq. \eqref{GrapheneConductivity1}] is reproduced. This requires that
\begin{align}\label{EffectiveMass0}
m_c &=- \frac{e^2}{2\pi c^2} \frac{|N_L-1/2|^2}{\chi_d}\\
&=\frac{\hbar k_F}{v_F}\left(1-\frac{1}{32|N_L-1/2|^2}\right)^{-1}
\,,
\label{EffectiveMass1}
\end{align}
where $k_F = \ell^{-1}\sqrt{2|N_L-1/2|}$ (per spin per valley) in view of which Eq. \eqref{EffectiveMass1} reduces to Eq.~(\ref{CyclotronMass}) -- the standard cyclotron mass in the limit of large $N_L$.
  
\subsection{\textit{Exact} evaluation of the orbital magnetic susceptibility}
\label{App:ExactChi}

Although derived for large $N_L$, Eq.~(\ref{EffectiveMass1}) is surprisingly accurate even for small $N_L$. This can be established by comparison with an exact evaluation of the orbital magnetic susceptibility, which can be done with the help of the Ramanujan formula for the sum of the square roots of the first $n$ natural integers \cite{Ramanujan1915}. According to Eq. (14) of Ref.~[\onlinecite{Ramanujan1915}] the sum of the square roots of the first $N$ natural integers is given by
\begin{align}\label{Ramanujan}
\sum_{k=1}^N \sqrt{k}=-C_1+\frac{2}{3}N^{3/2}+\frac{1}{2}N^{1/2}+\frac{1}{24 N^{1/2}} 
-\frac{1}{24}\sum_{k=0}^\infty\frac{1}{(\sqrt{N+k}+\sqrt{N+1+k})^5\sqrt{(N+k)(N+1+k)}}
\end{align}
where $C_1 = \frac{\zeta(3/2)}{4\pi}$ where $\zeta(z)$ is the Riemann zeta function (Note that $C_1$ appears with the wrong sign in Eq. (14) of Ref.~[\onlinecite{Ramanujan1915}].  We have used the correct sign here). The infinite series on the right of Eq.~(\ref{Ramanujan}) is rapidly converging and can be evaluated numerically without difficulty. Denoting by $F(N)$ the entire quantity on the right-hand side of Eq.~(\ref{Ramanujan}), we see that the energy density of Eq.~(\ref{EnergyDensity}) can be expressed as
\begin{align}\label{ExactEB} 
\epsilon(B)= -\frac{\hbar v_F\sqrt{2}}{2\pi\ell^3}\left[F(N_c)-F(N_L-1)\right]
\end{align} 
where $N_c = \frac{k_c^2\ell^2}{2}$ is the upper cutoff. Noting that the terms proportional to $N_c^{3/2}$ and $N_c^{1/2}$ do not contribute to the second derivative with respect to $B$, we discard them and then take the limit of $N_c \to \infty$ and find
\begin{align}
\epsilon(B)\simeq\frac{\hbar v_F\sqrt{2}}{2\pi\ell^3}\bigg[\frac{2}{3}(N_L-1)^{3/2}+\frac{1}{2}(N_L-1)^{1/2}&+\frac{1}{24 (N_L-1)^{1/2}}\nn\\ 
&-\frac{1}{24}\sum_{k=0}^\infty\frac{1}{(\sqrt{N_L-1+k}+\sqrt{N_L+k})^5\sqrt{(N_L-1+k)(N_L+k)}}\bigg],
\end{align}
where the $\simeq$ reminds us that we are excluding two formally infinite terms, which do not contribute to the orbital magnetic susceptibility. Taking the second derivative we arrive at
\begin{align}\label{ExactChid}
\chi_d=-\frac{3\sqrt{2}}{8\pi} \frac{e^2}{\hbar c} \frac {v_F\ell}{c} \bigg[\frac{2}{3}(N_L-1)^{3/2}+\frac{1}{2}(N_L-1)^{1/2}&+\frac{1}{24 (N_L-1)^{1/2}}\nn\\
&-\frac{1}{24}\sum_{k=0}^\infty\frac{1}{(\sqrt{N_L-1+k}+\sqrt{N_L+k})^5\sqrt{(N_L-1+k)(N_L+k)}}\bigg].
\end{align}
From Eq. \eqref{EffectiveMass0} and the expression for $k_F$, the effective cyclotron mass is then given by
\begin{align} \label{EffectiveMass2}
m_c = \frac{\hbar k_F}{v_F} \left(-\frac{1}{2\pi\sqrt{2}}\frac{e^2}{\hbar c}\frac{v_F\ell}{c}\frac{|N_L-1/2|^{3/2}}{\chi_d}\right)\,,
\end{align}
where the exact $\chi_d$ is given by Eq. \eqref{ExactChid}. 

In the special case $N_L=1$ from Eq. \eqref{ExactEB} we have
\begin{align}
\epsilon(B)=  -\frac{\hbar v_F\sqrt{2}}{2\pi\ell^3}F(N_c)\,,
\end{align} 
which, in the limit $N_c\to \infty$ leads to
\begin{align}\label{EBNL1}   
\epsilon(B) \simeq  \frac{\hbar v_F\sqrt{2}}{2\pi\ell^3}\frac{\zeta(3/2)}{4\pi}
\end{align} 
and
\begin{align}
\chi_d  = - \frac{3\sqrt{2}}{8\pi} \frac{e^2}{\hbar c} \frac {v_F\ell}{c}\frac{\zeta(3/2)}{4\pi}\,.
\end{align} 
It is noteworthy that the energy density in Eq. \eqref{EBNL1} has been calculated with different methods in earlier literature found in both high-energy (e.g. in the context of Landau levels for Dirac fermions) and condensed-matter physics (e.g. in the study of orbital antiferromagnetic state in square lattice for high-temperature superconductors) \cite{Chi_Refs}. 

Substituting this expression in Eq.~(\ref{EffectiveMass2}) with $N_L=1$ yields
\begin{align}\label{McNL1} 
m_c=\frac{\hbar k_F}{v_F}\frac{4\pi}{3\sqrt {2} \zeta(3/2)}.
\end{align} 

To compare the exact results with the approximate ones for the effective cyclotron mass, we note that Eq.~(\ref{EffectiveMass1}) yields $m_c=\frac{\hbar k_F}{v_F}\frac{8}{7}\simeq 1.143 \frac{\hbar k_F}{v_F}$ while the exact result from Eq. \eqref{McNL1} is $m_c\simeq 1.134 \frac{\hbar k_F}{v_F}$. It appears that the exact and the approximate results become practically indistinguishable for larger values of $N_L$.   

\twocolumngrid


\begin{thebibliography}{100}
\bibitem{Footnote1} 
	Restricting the attention to spatially uniform metric is justified by the fact that we are considering deformations of the electron liquid on a scale much larger than the lattice constant.  This assumption dramatically simplifies the Hamiltonian, which would otherwise contain additional terms related to the representation of the $\sigma$-matrices in a curved space time, as discussed in Appendix A of Ref.~[\onlinecite{Vozmediano10}].
\end{thebibliography}
\end{document}